\documentclass[final,3p,times]{elsarticle}

\usepackage{graphicx}
\usepackage{dcolumn}
\usepackage{bm}
\usepackage{graphics}
\usepackage{mathtools}

\usepackage{caption}
\usepackage{subcaption}

\usepackage{float}
\usepackage[normalem]{ulem}

\usepackage{hyperref}
\hypersetup{
    colorlinks,%
    citecolor=blue,%
    linkcolor=blue,%
    urlcolor=blue
}

\renewcommand{\eqref}[1]{(\ref{#1})}

\newcommand{\be}{\begin{equation}}
\newcommand{\ee}{  \end{equation}}
\newcommand{\ba}{\begin{eqnarray}}
\newcommand{\ea}{  \end{eqnarray}}

\graphicspath{ {./Figs/} }

\begin{document}

\title{Suppression of the Mott insulating phase in the particle-hole asymmetric Hubbard model}

\author{Mateus Marques $^1$, 
Bruno M. de Souza Melo $^2$,  
Alexandre R. Rocha $^3$, 
Caio Lewenkopf $^{2,4}$, 
Luis G.~G.~V. Dias da Silva $^1$}

\address{ $^1$ Instituto de F\'{\i}sica, Universidade de S\~ao Paulo, Rua do Mat\~ao 1371, S\~ao Paulo, S\~ao Paulo 05508-090, Brazil }

\address{$^2$ Instituto de F\'{\i}sica, Universidade Federal  Fluminense, Niter\'oi, Rio de Janeiro 24210-346, Brazil}

\address{$^3$ Instituto de F\'{\i}sica Te\'orica, S\~ao Paulo State University (UNESP), S\~ao Paulo, S\~ao Paulo 01140-070, Brazil}

\address{$^4$ Instituto de F\'{\i}sica, Universidade Federal  do Rio de Janeiro, Rio de Janeiro, Rio de Janeiro 21941-909, Brazil}

\begin{abstract}
We explore the phase diagram of the Mott metal-insulator transition (MIT), focusing on the effects of particle-hole asymmetry (PHA) in the single-band Hubbard model. Our dynamical mean-field theory (DMFT) study reveals that the introduction of PHA in the model significantly influences the critical temperature ($T_c$) and interaction strength ($U_c$), as well as the size of the co-existence region of metallic and insulating phases at low temperatures. Specifically, as the system 
is moved away from particle-hole symmetry, $T_c$ decreases and $U_c$ increases, indicating a suppression of the insulating phase and the strengthening of the metallic behavior. Additionally, the first-order transition line between metallic and insulating phases is better defined in the model with PHA, leading to a reduced co-existence region at $T<T_c$. Moreover, we propose that the MIT can be characterized by the charge density, which serves as a viable alternative to zero-frequency spectral density typically used in DMFT calculations. Our findings provide new insights into the role of particle-hole asymmetry in the qualitative and quantitative characterization of the MIT even in a 
very simple system. 
\end{abstract}


\maketitle

\section{Introduction}
\label{sec:introduction}

A complete understanding of strong correlations in materials remains one of the greatest challenges in condensed matter physics \cite{Peters2009, Haule2010}.
The competition between itinerant and correlated behavior of the electrons gives rise to rich and complex phenomena with promising applications such as oxide electronics, high-temperature superconductors, and spintronic devices \cite{Choi2016, Ruegg2014, Zhou2013, Zitko2009}.

A classical topic in this field is the study of the Mott metal-insulator transition (MIT) \cite{Mott1968}, occurring  when the electron-electron interactions are relatively large, of the order of the bandwidth. 
In this scenario, the electrons can become localized, 
leading to an insulating state.
On the other hand, as the interactions are reduced and the electrons can become itinerant, the system 
transitions into
a strongly correlated metal phase \cite{Kajueter1996}. In actual experiments, the MIT can be induced by varying temperature, pressure, and composition of the materials in the system \cite{Sayyad2019,Qiu2017,Rozenberg1994a}.

One conventional approach to study the qualitative effects of strong correlations is the proposition and solution of model Hamiltonians. These models drastically reduce the number of degrees of freedom of the problem to the relevant ones involved in a given phenomenon. From a theoretical point of view, the Hubbard model \cite{Hubbard1963,Hubbard1964} has been successfully used to
describe the main features of Mott insulators, like the transition metal oxide $\text{V}_2 \text{O}_3$ \cite{Rozenberg1994a, Kotliar2002}.

The MIT has been extensively studied in the literature, mostly using the dynamical mean field theory (DMFT) approach \cite{Georges1996,Held2007}  in models with particle-hole symmetry (PHS). In this case, it is well established that, below a critical temperature, 
there is a first-order transition line in the temperature ($T$) versus interaction ($U$) phase diagram ending in the critical point $(U_c, T_c)$. For $T<T_c$,
the system is metallic for $U<U_{c_1}(T)$ and insulating for $U>U_{c_2}(T)$ with 
a coexistence region (where one of the phases is thermodynamically stable while the other is only metastable)
for values of $U$ between these two values \cite{Georges1996,Vildosola2015,Bulla2001, Joo2001, Eckstein2010,Jarrell1993a}. The two 
spinodal lines defined by $U_{c_1}(T)$ and $U_{c_2}(T)$  meet at the critical point $(U_c, T_c)$. For $T>T_c$, the transition becomes a crossover \cite{Rozenberg1994a,Kotliar2002,Tong2001}.  

Interestingly, most of the DMFT investigations have concentrated on the low-temperature regime ($T<T_c$), while the high-temperature crossover region 
is the one
pertinent to numerous experimental setups \cite{Terletska_PRL.107.026401_2011}. Conversely, for $T>T_c$, different crossover regimes have been tentatively discerned \cite{Georges1996}, yet they have not been examined in any significant depth \cite{Dobrosavljevic_Book}.

Although less studied, the particle-hole asymmetric (PHA) Hubbard model is of particular interest since 
that is the prevalent situation for the
band structure \cite{Vildosola2015, Demchenko2004,Eckstein2007}
of most materials. 
Particle-hole asymmetry can be implemented 
in the Hamiltonian model
by including next-nearest neighbor hopping 
matrix elements 
at half-filling or by explicitly changing the onsite energy \cite{Ruegg2014, Kajueter1996, Kotliar2002, Eckstein2007, Garcia:Phys.Rev.B:121102:2007, Fisher1995,Werner2007}. As argued above, although the MIT in the Hubbard model has been extensively studied, most of the studies focus on the particle-hole symmetric (PHS) case as this naturally renders the system to be in half-filling (that corresponds to an occupation number $n=1$) independently of the interaction strength $U$. 
A crucial fact is that the MIT can occur \emph{away from PHS} and this will be marked by $n \rightarrow 1$ at the transition \cite{Garcia:Phys.Rev.B:121102:2007,Werner2007}. In other words, at the insulating side of the MIT, the system will be both particle-hole asymmetric (PHA) \emph{and} at half-filling at the same time.

In this work we revisit this problem by studying the Mott metal-insulator transition (MIT) in the particle-hole asymmetric Hubbard model in the Bethe lattice.  While earlier studies have addressed the transition away from PHS at zero  \cite{Garcia:Phys.Rev.B:121102:2007,Logan:J.Phys.Condens.Matter:025601:2015} or very low temperatures \cite{Werner2007}, here we show that the PHA system has a rich phase diagram at non-zero temperatures. 
As such, our focus here is on the fate of the co-existence region \emph{away} from PHS at larger temperatures. In fact, our results show that this region is \emph{suppressed} as the system moves away from PHS, a fact which has not been properly documented in previous works. 

This behavior is characterized by a lower value of the maximum temperature $T_c$ for which a first-order transition between the metallic and insulating phases 
takes place. 
For $T>T_c$, the transition is second-order, such that $T_c$ marks the ``end'' of the first-order transition line \cite{Terletska_PRL.107.026401_2011}. As such, this critical temperature $T_c$ \emph{decreases} as the system moves away from PHS.
Although similar decreases in $T_c$ have been seen in doped \cite{Vucicevic:Phys.Rev.Lett.:246402:2015} and disordered systems \cite{Carol_PhysRevB.73.115117,Carol_PhysRevB.92.125143}, here we provide a systematic analysis of these effects in a ``clean'' system in which the on-site energy can be independently varied (say, by an external gate-voltage). As such, our approach can be extended to systems in which the density $n$ is not fixed \emph{a priori} and can vary across the MI transition.

More importantly, in the realistic case of systems away from PHS, we show that the charge density can be a reliable marker of the true MIT transition not only at $T \sim 0$ \cite{Werner2007} but also for larger temperatures, introducing the interesting prospect of experimentally characterizing the MIT by the electron (or hole) densities. This allows one to characterize the transition by the charge density $n$ rather than by the zero-frequency spectral density $\rho(0)$, the usual parameter considered in theoretical works.

This paper is organized as follows. In Sec.~\ref{sec:model} we briefly present the model and methods used in this study. In Sec.~\ref{sec:results} we show results for the symmetric and asymmetric Hubbard model focusing in the differences displayed by the two cases with regard to the metal-insulator transition. Finally, in Sec.~\ref{sec:conclusion} we
present a summary of our findings and conclusions.

\section{Model and methods}
\label{sec:model}

The single-band Hubbard model \cite{Hubbard1963} reads
\begin{equation}
\hat{H} = - t \sum_{\langle ij \rangle, \sigma} c_{i \sigma }^{\dagger} c^{}_{j \sigma} + {\rm H.c.}  + U \sum_{i } n_{i \uparrow} n_{i \downarrow} -\epsilon_d \sum_{i, \sigma} n_{i \sigma} \; ,
\label{eq:Hubbard_model}
\end{equation}
where $c_{j \sigma}^{\dagger} (c^{}_{j \sigma})$ is the creation (annihilation) operator of an electron with spin $\sigma$ at site $j$, $n_{j \sigma} = c_{j \sigma}^{\dagger} c_{j \sigma}$ is the occupation number
operator, $t$ is the hopping amplitude, and $\langle \cdots \rangle$ means that 
the sum is restricted to nearest neighbors sites.

Our study focuses on the electronic spectral function $\rho(\omega)$ and on the single-electron density $n$ given by
\begin{equation}
n = \int_{-\infty}^{\infty} f(\omega,T) \rho(\omega) d \omega \; ,
\end{equation}
where $f(\omega,T) = (e^{\omega/k_B T} + 1)^{-1}$ is the Fermi-Dirac function at a temperature $T$. 
Hence, our goal is to calculate the local lattice interacting single-particle Green's function.

Given the ${\bf k}$-dependent lattice Green's function $G({\bf k}, \omega)$, we use the dynamical mean field theory (DMFT) approach \cite{Georges1996,Held2007} to obtain the local single-particle Green's function in momentum space $G_{\text{loc}}(\omega)= \sum_{{\bf k}} G({\bf k}, \omega)$ and its spectral function $\rho(\omega)=(-1/\pi) \, \text{Im} \, \{G_{\text{loc}}(\omega)\}$. 
In essence, DMFT maps a lattice many-body problem with many degrees of freedom into an effective single-impurity problem within a self-consistent cycle. As such, this method addresses the limiting and intermediate regimes of the ratio $U/t$ of the model in a single framework. 

For the Hubbard model the effective problem can be treated using the single impurity Anderson model (SIAM) \cite{Anderson1961, Georges1996, Kotliar2006}.
The implementation of DMFT for different settings is nicely reviewed in Refs.~\cite{Georges1996,Kotliar2006, Held2007}.

We consider the Hubbard model in a Bethe lattice in the limit of infinite coordination ($z \to \infty$). 
In this case, the hopping needs to be renormalized as $t = t_*/\sqrt{z}$ for the limit to make physical sense. The non-interacting density of states (DOS) then reads \cite{Georges1996,Economou2006}
\begin{equation}
 \rho_{0}(\epsilon) = \frac{1}{2 \pi t_*^2}\sqrt{4t_*^2-\epsilon^2}, \quad |\epsilon|<2t_*,
\label{eq:DOS_Hypercubic_Lattice}
\end{equation}
where $D = 2t_*$ is the half-bandwidth of the non-interacting model, which we use as our energy unit.

In the effective impurity approximation \cite{Georges1996}, the single site electrons
are coupled to the non-interacting bath through an energy-dependent  hybridization function whose form depends on the 
lattice effective degrees of freedom.
For a Bethe lattice, the hybridization function takes the simple form \cite{Georges1996}:
\begin{equation}
\Delta(\omega)=t^2G_{\rm imp}(\omega)~,
\end{equation}
which is calculated self-consistently from the single-site (impurity) Green's function 
$G_{\rm imp}(\omega)=(\omega+\epsilon_d-\Delta(\omega)-\Sigma^{I}(\omega))^{-1}$
where 
$\Sigma^{I}(\omega)$ is the interacting self-energy.

The Green's function is obtained by solving the auxiliary impurity model. 
Over the last decades, several different impurity solvers have been developed like the iterated perturbation theory (IPT) \cite{Yosida1970,Yosida1970a,Yamada1975}, numerical renormalization group (NRG) \cite{Wilson1975,Bulla1999}, equations of motion (EOM) schemes \cite{Theumann1969, Lacroix1981, Lacroix1982, Meir1991, Kashcheyevs2006, VanRoermund2010, Lavagna2015}, auxiliary-boson approaches \cite{Coleman1984}, including the non-crossing (NCA)\cite{bickers1987} and one-crossing (OCA) approximations \cite{Pruschke1989}, quantum Monte Carlo (QMC) \cite{Hirsch1986}. 
For each method, there is a region in the parameter space for which the results are more accurate. 
NRG, for instance, provides excellent real-frequency spectral data at zero temperature, while EOM and the NCA methods work better at higher temperatures. 
NCA and EOM, however, have well-known limitations in establishing quantitative estimates for quantities such as the critical temperature  $T_c$ as compared to, say, QMC \cite{Dobrosavljevic_Book}. 
The computational cost is also an important aspect to be considered as NRG and QMC are computationally much more demanding than the EOM or auxiliary-boson approaches \cite{Melo2020}.

In this work, we use the non-crossing approximation (NCA)  \cite{bickers1987,Sposetti2016} as an impurity solver \footnote{For the calculations, we used the NCA code from Kristjan Haule available at \url{http://hauleweb.rutgers.edu/tutorials/} }. 
While NCA has well-documented limitations, like in accurately quantifying the parameters $U_{c_1}$ and $U_{c_2}$ \cite{Vildosola2015}, it remains a valuable tool for its low computational cost and ability to incorporate finite temperatures \cite{Melo2020}.
These features allow us to quickly assess a large range of parameters to get a qualitative insight of the main features and differences between the particle-hole symmetric and asymmetric models \cite{Bulla2001,Grewe2008} at $T > 0$.

\section{MIT in the Hubbard model away from PHS}
\label{sec:results}

\subsection{Signatures of the MIT  from the DMFT results }
\label{sec:mittc}

\begin{figure}[h!]
\centering
\includegraphics[width=0.8\columnwidth]{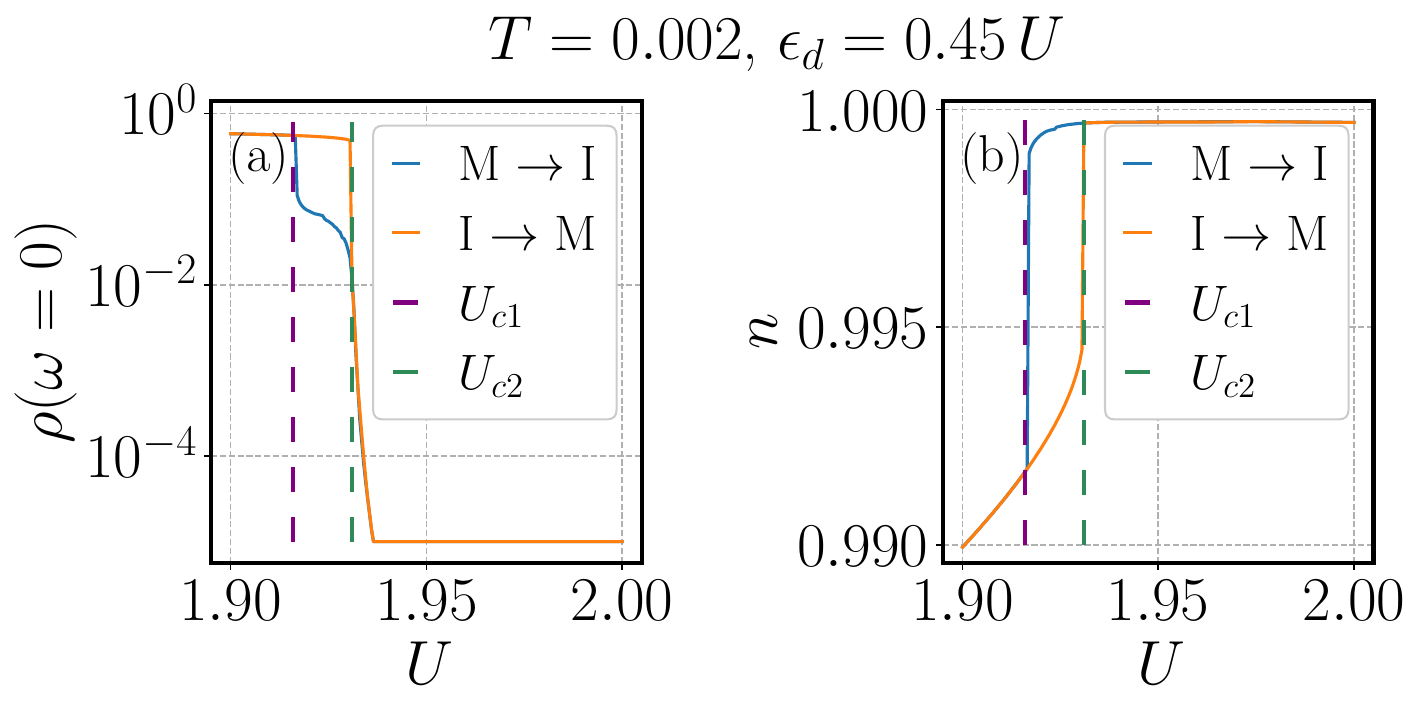}
\includegraphics[width=0.8\columnwidth]{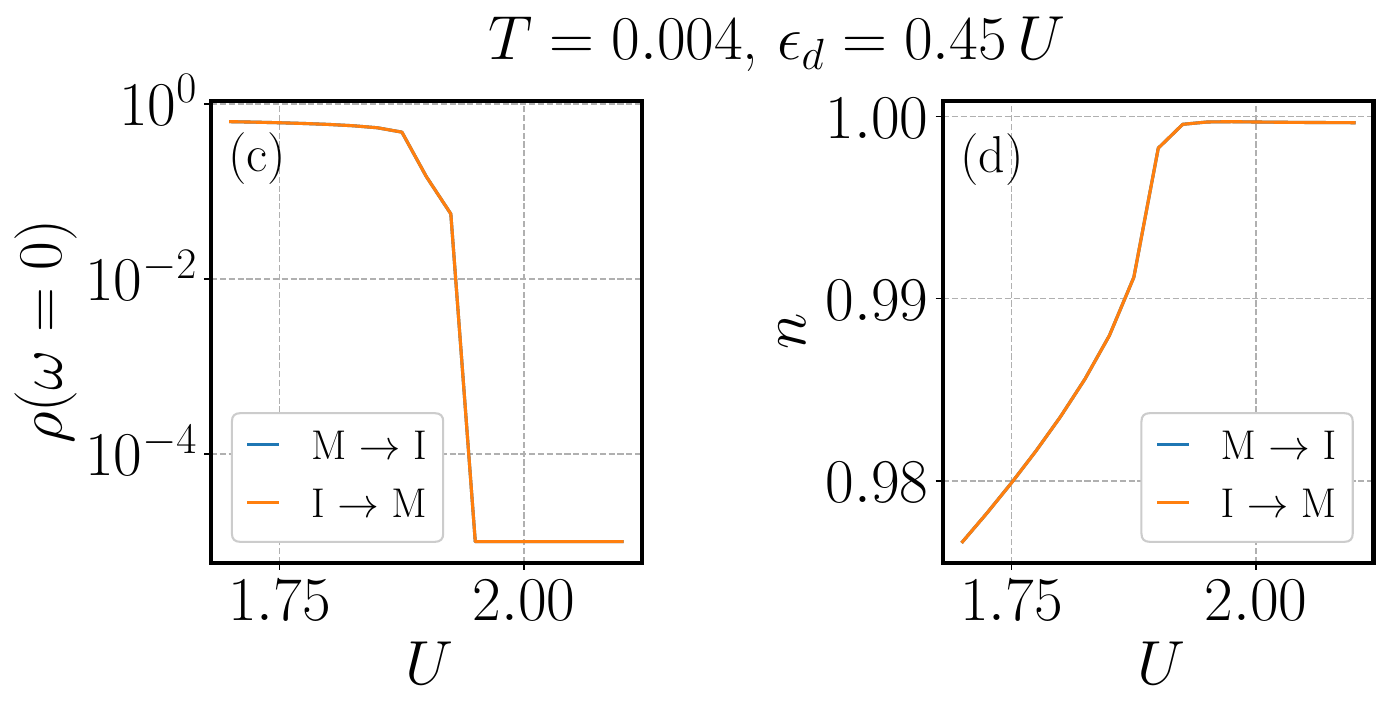}
\caption{Occupation $n$ and zero-frequency spectral density $\rho(0)$ 
versus $U$ for the metal-insulator M $\rightarrow$ I and the insulator-metal I $\rightarrow$ M paths for temperatures below and above $T_c$. Both paths coincide for $T>T_c$.
}
\label{fig:n_rho_vsU_T0002_T0004_ed045}
\end{figure}

When the system is away from the particle-hole symmetric point, 
$\epsilon_d \neq 0.5 \, U$, the MIT transition can be characterized by either the 
occupation
$n$ or the zero-frequency spectral density $\rho(0)$ as a function of $U$. 
This is illustrated in Figure \ref{fig:n_rho_vsU_T0002_T0004_ed045}, which 
displays 
$n$ and $\rho(0)$ as a function of $U$ for $\epsilon_d=0.45 \, U$ 
at two distinct temperatures 
(below and above $T_c$).

As it is well-known \cite{Georges1996,Dobrosavljevic_Book}, for temperatures lower than a ``critical temperature'' ($T<T_c$), a co-existence region occurs for $U_{c_1} < U < U_{c_2}$, with first-order transitions occurring at $U=U_{c_1}$ or $U=U_{c_2}$ depending on the path chosen ($\text{M} \to \text{I}$ or $\text{I} \to \text{M}$, respectively), forming a 
``hysteresis loop''.

\begin{figure}[t!]
\centering
\includegraphics[width=0.8\columnwidth]{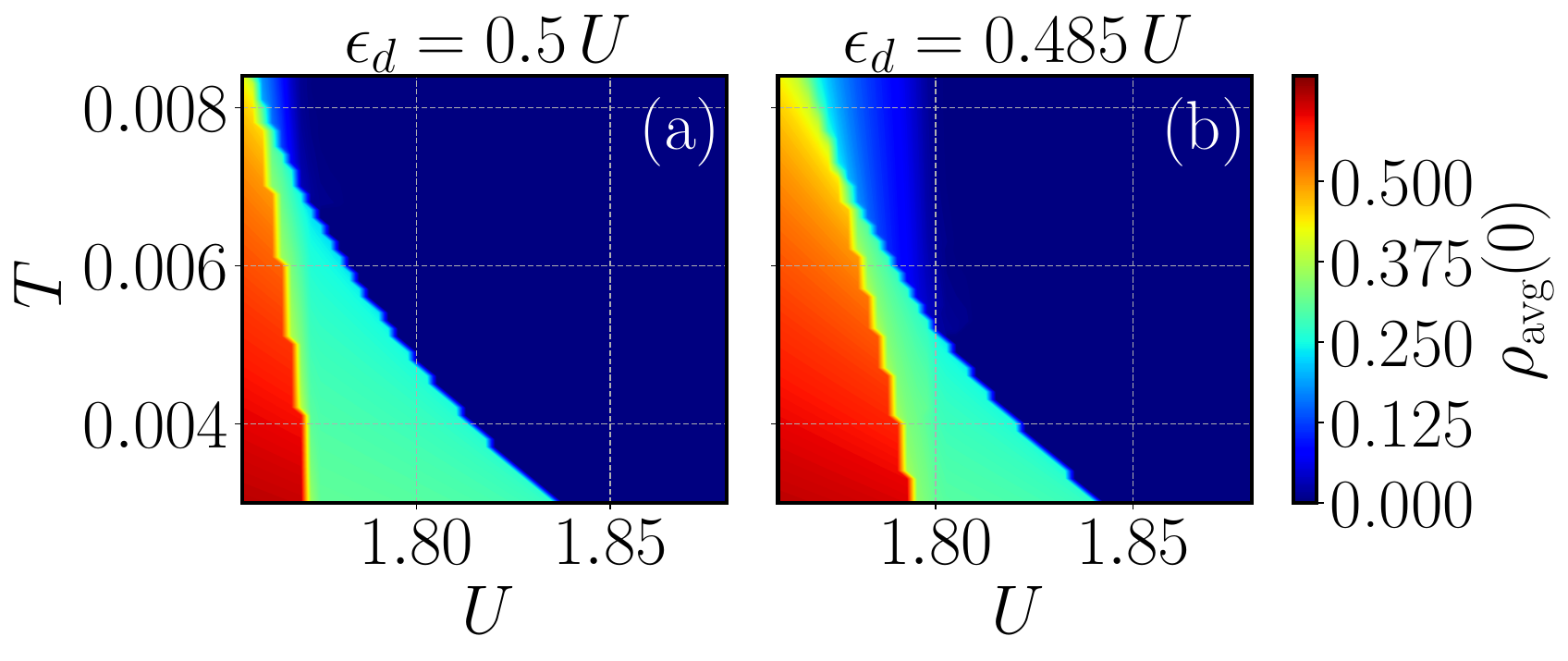}
\includegraphics[width=0.8\columnwidth]{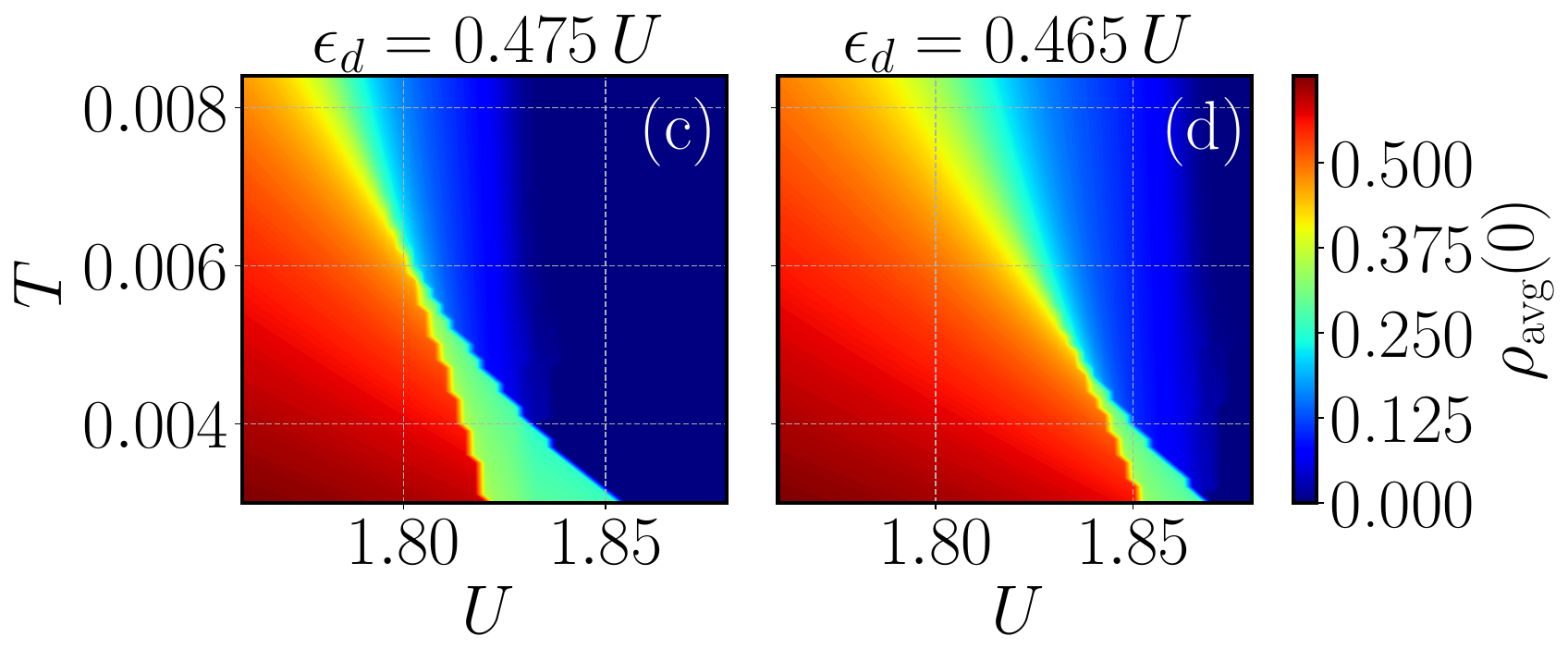}
\includegraphics[width=0.8\columnwidth]{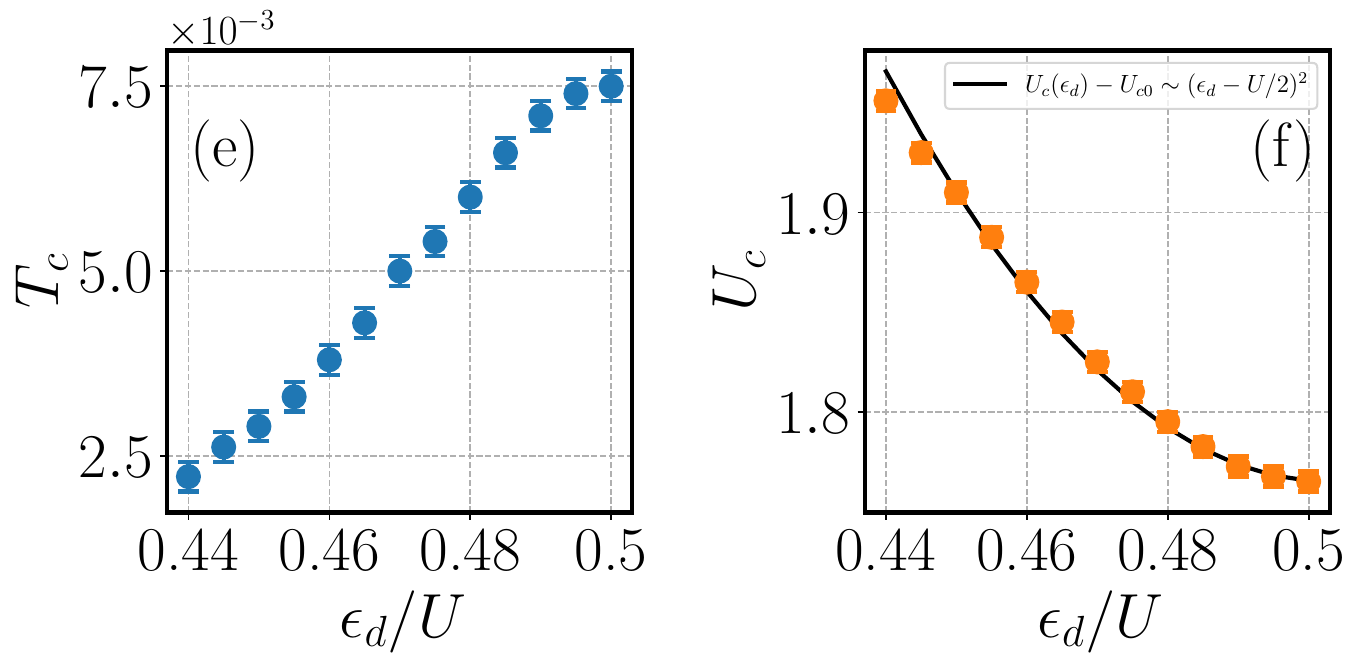}
\vspace{-1ex}
\caption{ (a-d) Phase diagrams  $U \times T$ for the Hubbard model as it is moved away from particle-hole symmetry: (a) $\epsilon_d=0.5U$ (PHS) (b) $\epsilon_d=0.485U$ (c) $\epsilon_d=0.475U$ (d) $\epsilon_d=0.465U$. 
Notice that the co-existence region becomes significantly smaller as 
the system is moved away from the PHS point. 
(e-f) $T_c$ and $U_c$ (obtained from the method described in the main text) versus $\epsilon_d$, showing a decrease in $T_c$ and an increase in $U_c$ as 
$\epsilon_d$ is shifted away from the PHS point.
}
\label{fig:TvsU_ed_Tcvsed}
\end{figure}

These features are clearly identified
in the DMFT data shown in Figure \ref{fig:n_rho_vsU_T0002_T0004_ed045} for lower temperatures. 
At $T = 0.002$, both $\rho(0)$ and $n$ exhibit first-order phase transitions at different values of $U$, depending on the path taken, as illustrated in Figs.~\ref{fig:n_rho_vsU_T0002_T0004_ed045}(a) and (b).
From these results, we can extract the values of $U_{c_1}$ and $U_{c_2}$, which are marked in the plots. 
On the other hand, 
at $T=0.004$,
the co-existence region is negligible and $U_{c_1}=U_{c_2} \equiv U_c$,
as shown by Figs.~\ref{fig:n_rho_vsU_T0002_T0004_ed045}(c) and (d).

As a standard procedure, it is useful to define the following quantities:
\begin{align}
\label{eq:rhoavgrhodiff}
    \rho_{\rm avg}(0) \equiv &\, \left( \rho_{\text{M} \rightarrow \text{I}}(0) + \rho_{\text{I} \rightarrow \text{M}}(0) \right)/2 \; , 
    \\ \nonumber
    \Delta \rho(0) \equiv &\,|\rho_{\text{I} \rightarrow \text{M}}(0) - \rho_{\text{M} \rightarrow \text{I}}(0)| \; , 
\end{align} 
where $\rho_{\text{M} \rightarrow \text{I}}(0)$ and $\rho_{\text{I} \rightarrow \text{M}}(0)$ are calculated in the $\text{M} \to \text{I}$ and $\text{I} \to \text{M}$ paths respectively.

From these 
quantities, we can numerically obtain the transition temperature $T_c$ and the values of $U_{c_1}$ and $U_{c_2}$ by the following method: 
For each temperature $T$ and path ($\text{M} \to \text{I}$ or $\text{I} \to \text{M}$), we take the numerical derivative of either $\rho(0)$ 
or
$n$ with respect to $U$. 
These derivatives 
show peaks at the transition points, which 
correspond to
the values of $U_{c_1}$ and $U_{c_2}$, depending on the path. 
We can then obtain $T_c$ by 
identifying
the value of $T$ for which these two peaks ``merge'' (such that $U_{c_1} \approx U_{c_2}$). The numerical error associated with the value of $T_c$ obtained in this way can be reduced by decreasing the discretization steps in both $T$ and $U$.

\subsection{Phase diagrams away from PHS }
\label{sec:phacase}

From these considerations, one can then represent the phase diagram of the system by color plots for $\rho_{\rm avg}(0)$ as a function of $T$ and $U$. These are shown in Figure \ref{fig:TvsU_ed_Tcvsed} (a)-(d)  for different values of the onsite energy $\epsilon_d$. These  define $T$ versus $U$ phase diagrams showing the insulating (blue), metallic (red) and co-existence (green) regions. As expected \cite{Georges1996,Dobrosavljevic_Book}, the difference between $U_{c_2}$ and $U_{c_1}$ decreases as $T$ increases, such that the co-existence regions narrows to a single point at a critical temperature $T_c$ at which $U_{c_2}=U_{c_1} = U_c$ .

\begin{figure}[t!]
\includegraphics[width=0.8\columnwidth]{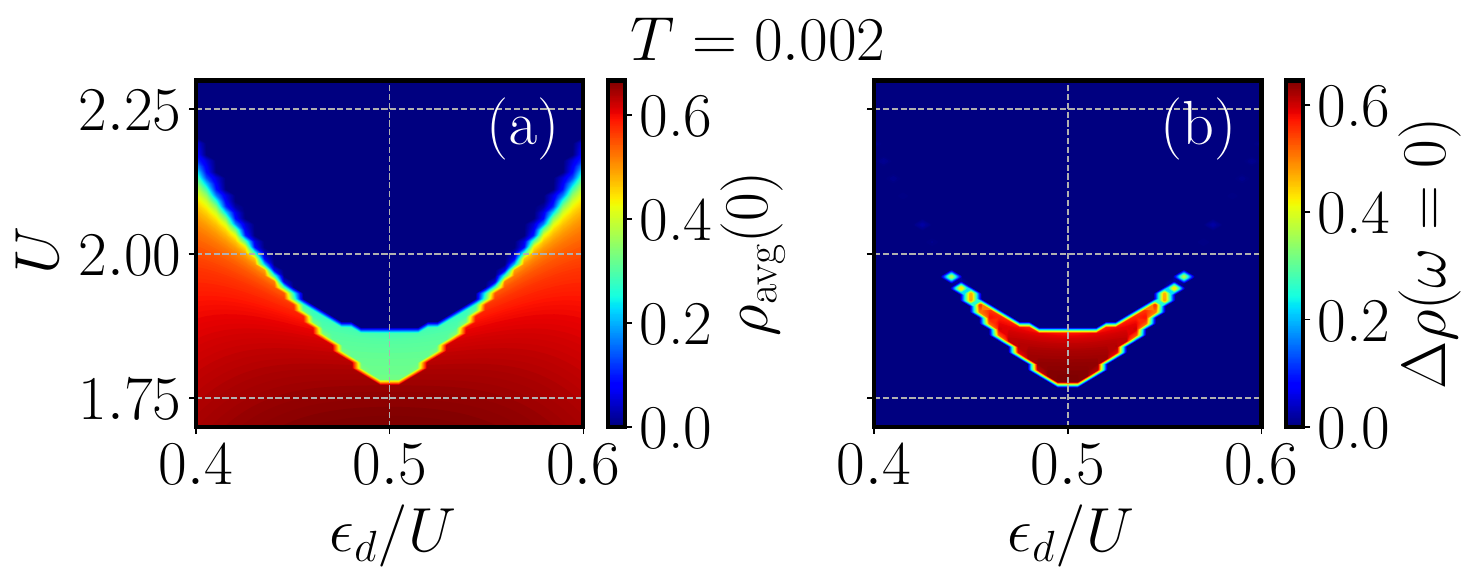}
\includegraphics[width=0.8\columnwidth]{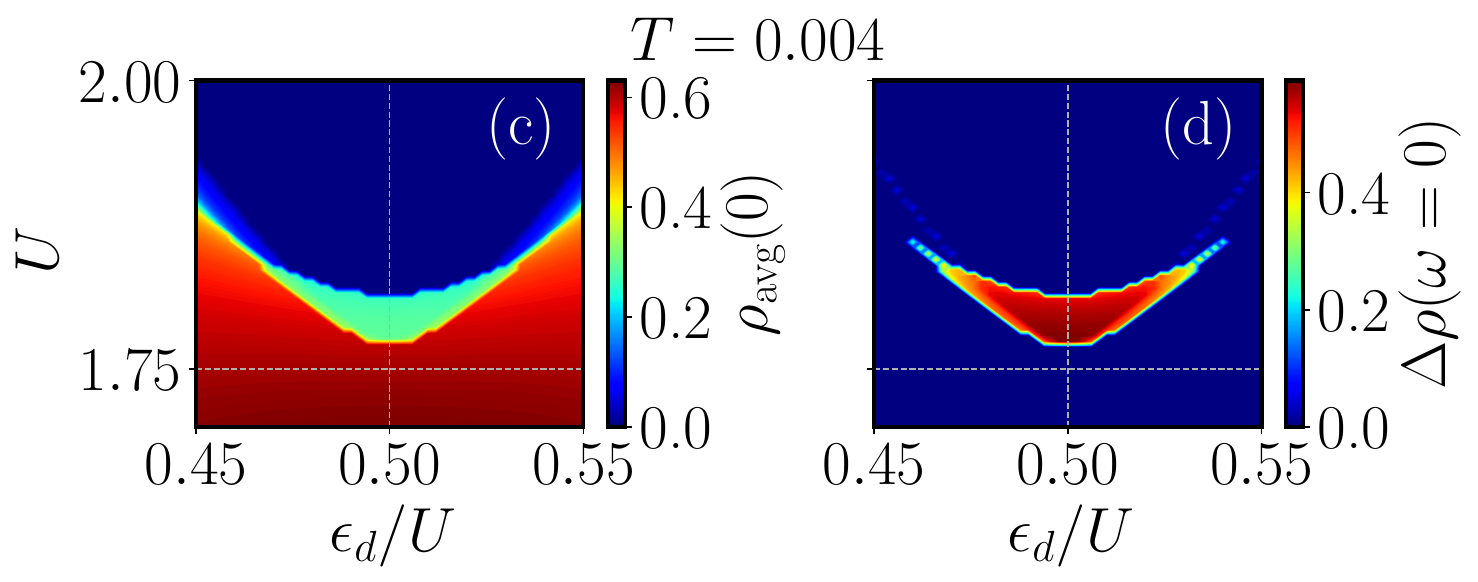}
\caption{
DMFT-NCA phase diagram $U \times \epsilon_d$ 
of
the Hubbard model for $T=0.002$ (a,b) and $T=0.004$ (c,d). Panels (a,c) show $\rho_{\rm avg}(0)$ while panels (b,d) shows $\Delta \rho (0)$. }
\label{fig:Diagram_Drho_U_ed_T0002_T0004}
\end{figure}

Interestingly, the value of $T_c$ (and the overall area of the coexistence region) \emph{decreases} as the system is moved
away from the particle-hole symmetric point ($\epsilon_d=0.5 \, U$). 
This is illustrated in Fig.~\ref{fig:TvsU_ed_Tcvsed}(e), which shows a sharp decrease in $T_c$ as $\epsilon_d$ is reduced from $0.5 U$ (PHS point) to $0.44 U$
\footnote{Numerical instabilities in the NCA calculations prevented us from exploring the regime of temperatures lower than $T\sim 0.002$, which restricts our analysis to the range 
$0.44 U <\sim \epsilon_d \leq 0.5 U$
}. 
In addition, the corresponding value of $U_c$ \emph{increases} as $\epsilon_d$ is reduced in the same range, as depicted in Fig.~\ref{fig:TvsU_ed_Tcvsed}(f). In fact, the $U_{c}$ vs $\epsilon_{d}$ curve is very well approximated by a quadratic scaling of the form $U_{c}(\epsilon_{d}) - U_{c0} \sim (\epsilon_{d} - U/2)^2$ (solid line in Fig.~\ref{fig:TvsU_ed_Tcvsed}(f)), where  $U_{c0}$ is the minimal value at the particle-hole symmetry point $\epsilon_{d}=U/2$. This excellent agreement with a quadratic scaling is consistent with  analytical expressions relating $U_c$ and the chemical potential in the context of the doping-driven Mott transition \cite{Werner2007,Logan:J.Phys.Condens.Matter:025601:2015}.

Alternatively, phase diagrams can be visualized by plotting color maps of $\rho_{\rm avg}(0)$ and $\Delta \rho(0)$, defined in Eq.~\eqref{eq:rhoavgrhodiff}, in the $U \times \epsilon_d$ plane for a given temperature $T$. 
This is illustrated for $T = 0.002$ in Figs.~\ref{fig:Diagram_Drho_U_ed_T0002_T0004}(a) and (b) and for $T = 0.004$ by  Figs.~\ref{fig:Diagram_Drho_U_ed_T0002_T0004}(c) and (d).
We note that both $\rho_{\rm avg}(0)$ and $\Delta \rho(0)$ are symmetric with respect to the PHS value $\epsilon_{d}=0.5U$ for all temperatures. Consequently, the curves $U_{c}(\epsilon_{d})$ and $T_{c}(\epsilon_{d})$ shown in Fig.~\ref{fig:TvsU_ed_Tcvsed} are also symmetric with respect to $\epsilon_{d}=0.5U$.
These plots show more clearly the narrowing of the co-existence region,
corresponding to the green area in Figs.~\ref{fig:Diagram_Drho_U_ed_T0002_T0004}(a) and (c), 
and the red region in Figs.~\ref{fig:Diagram_Drho_U_ed_T0002_T0004}(b) and (d), as the system is moved away from the PHS point. 

\begin{figure}[t!]
\includegraphics[width=\columnwidth]{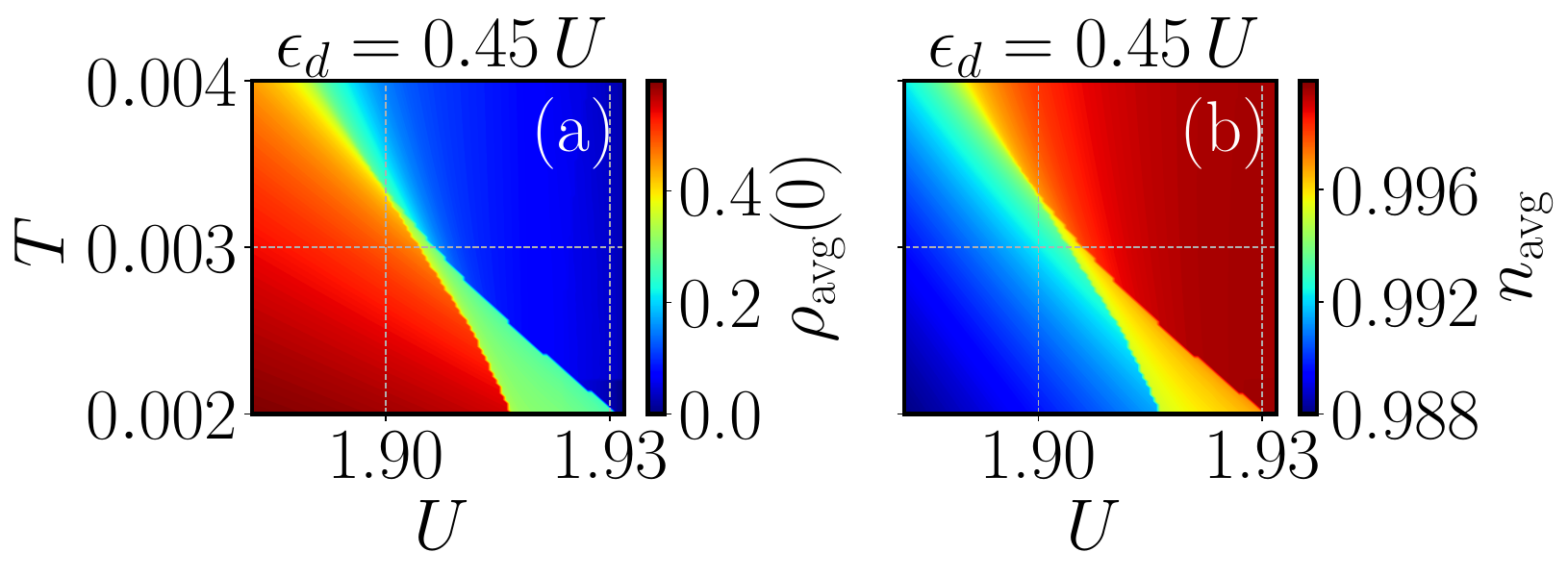}
\caption{
DMFT-NCA phase diagram  $U \times T$ for the Hubbard model away from particle-hole symmetry ($\epsilon_d=0.45U$). 
Panel (a) shows the average $\rho_{\rm avg}(0)$, while panel (b) shows the average density $n_{\rm avg}$, which is remarkably similar.}
\label{fig:Diagram_UxT_ed045}
\end{figure}

\subsection{Signatures of the MIT in the charge density }
\label{sec:chargedensity}

An important point regarding the phase diagrams away from the PHS is that different phases in the system can be characterized not only by the zero-frequency spectral density $\rho(0)$ but also by the \emph{charge density} $n$ since the insulating region is characterized by half-filling, namely, $n=1$. 
As such, the $T$ versus $U$ phase diagrams can be plotted by looking at $n$ alone. In an analogy with $\rho_{\rm avg}(0)$ and $\Delta \rho(0)$ defined in Eq.~\eqref{eq:rhoavgrhodiff}, it is useful to define the following quantities related to the charge density:
\begin{align}
\label{eq:navgrhodiff}
    n_{\rm avg} \equiv &\, \left( n_{\text{M} \rightarrow \text{I}} + n_{\text{I} \rightarrow \text{M}} \right)/2 \; , 
    \\ \nonumber
    \Delta n  \equiv &\, |n_{\text{I} \rightarrow \text{M}}- n_{\text{M} \rightarrow \text{I}}| \; , 
\end{align} 
where $n_{\text{M} \rightarrow \text{I}}$ and $n_{\text{I} \rightarrow \text{M}}$ are calculated in the $\text{M} \to \text{I}$ and $\text{I} \to \text{M}$ paths respectively.

\begin{figure}[t!]
\includegraphics[width=\columnwidth]{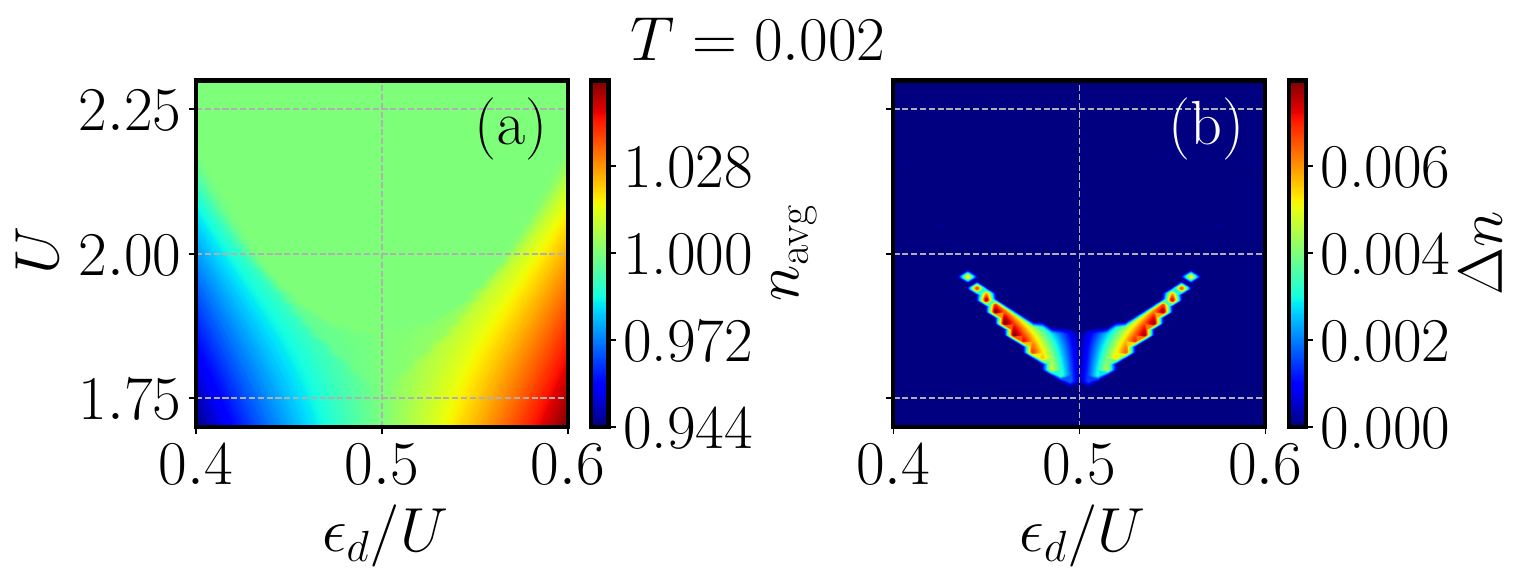}
\includegraphics[width=\columnwidth]{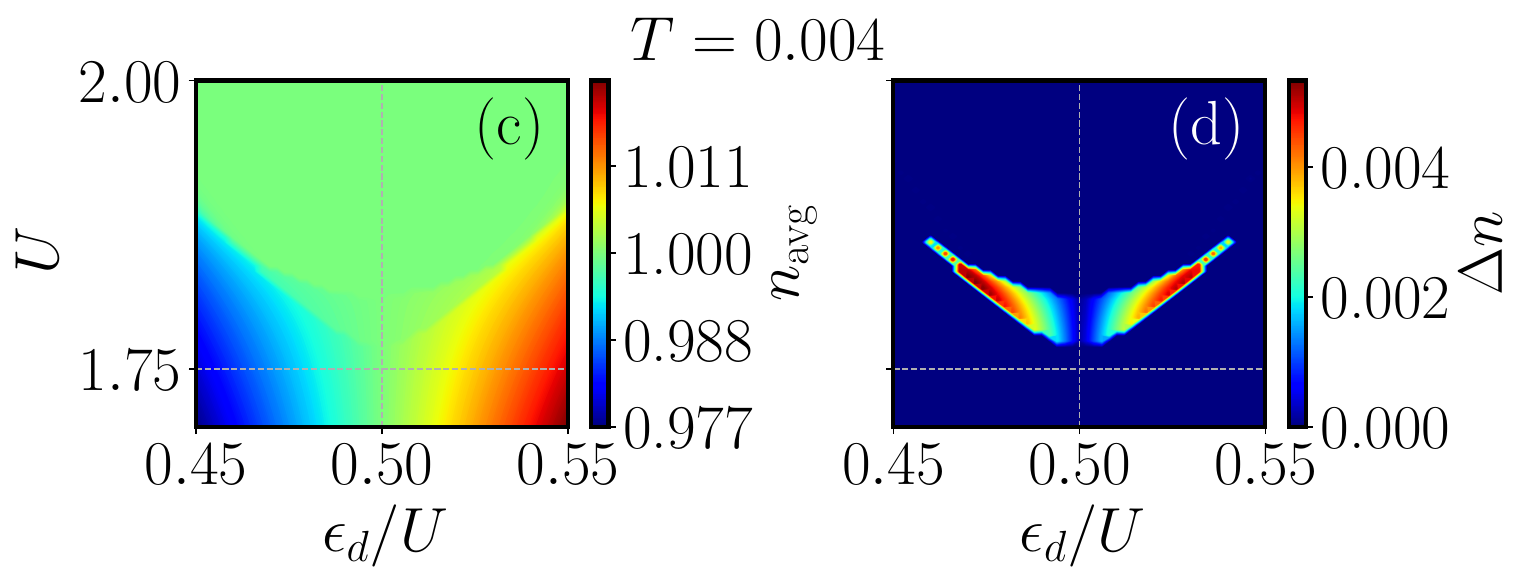}
\caption{Phase diagram  $U \times \epsilon_d$ for the Hubbard model for $T=0.002$ (a,b) and $T=0.004$ (c,d). Panels (a,c) show $n_{\rm avg}$ while panels (b,d) shows $\Delta n$. }
\label{fig:Diagram_n_U_ed_T0002_T0004}
\end{figure}

This one-to-one correspondence 
between the two quantities is evident
in Figs.~\ref{fig:Diagram_UxT_ed045}(a) and (b), which present 
similar $U \times T$ phase diagrams 
obtained from
$\rho_{\rm avg}(0)$ and $n_{\rm avg}$, respectively, for $e_d=0.45U$. Consequently, the phase diagrams of the system can also be effectively depicted by utilizing 
$n_{\rm avg}$. 
These 
are
displayed in Figs~\ref{fig:Diagram_n_U_ed_T0002_T0004} (a) to (d), which show 
$n_{\rm avg}(\epsilon_d,U)$ and $\Delta n (\epsilon_d,U)$ for $T=0.002$, see Figs.~\ref{fig:Diagram_n_U_ed_T0002_T0004}(a,b), and $T=0.004$,  see Figs.~\ref{fig:Diagram_n_U_ed_T0002_T0004}(c,d). 

The goal here is to emphasize that the density $n$, an experimentally accessible quantity, can be used to effectively characterize the MIT transition away from the PHS point. Of course, this does not hold at the PHS point ($\epsilon_d=0.5U$), since the system is always at half-filling ($n=1$) independently of the phase.

\section{Summary and conclusions}
\label{sec:conclusion}

In this work, we have revisited the Mott metal-insulator transition (MIT) within the framework of the particle-hole asymmetric (PHA) Hubbard model on the Bethe lattice. 
For a temperature below a critical value, a phase transition takes place for two values of interaction, $U_{c_1}$ and $U_{c_2}$, with a coexistence region between 
the latter.
Above a critical temperature there is a crossover between the metal and insulator phases. However, for the asymmetric model we found a first-order transition only for $U_{c_1}$. For $U_{c_2}$ we found a crossover between the two solutions with a pattern almost independent on the temperature.

Our results show that, in 
contrast with the well-studied particle-hole symmetric case, the PHA model exhibits a notably complex $U \times T$ phase diagram showing reduced value of the critical temperature $T_c$ as well as a narrower co-existence region between metallic and insulating phases at low temperature.
 This is illustrated in the $U \times T$ and $U \times \epsilon_d$ phase diagrams, indicating the ``shrinking'' co-existence region with increasing temperature and/or breaking the particle-hole symmetry.
 The plots indicate that the transition between metallic and insulating phases is better defined when the system is away from PHS. 
 
 Away from the particle-hole symmetric point, the MIT is characterized by a co-existence region where the system exhibits both metallic and insulating properties over a range of interaction strengths $U$. This region, marked by ``hysteresis loops'' in the density and spectral density plots, becomes more pronounced at lower temperatures and is strongly influenced by the departure from PHS. This suggests that the first-order transition line is more sharply defined in the asymmetric model, offering clearer distinctions between the two phases.

Moreover, a key contribution of our work is the identification of the charge density $n$ as a robust marker for the MIT transition in systems where particle-hole symmetry is broken since the transition will be marked by $n \rightarrow 1$. 
We notice that this behavior should occur in
realistic material systems
where PHS is broken by an asymmetric density of states,
which can be schematically modeled, for instance, by
a Bethe lattice
incorporating
second-neighbor hopping \cite{Eckstein2005}. 
Unlike traditional approaches that rely on the zero-frequency spectral density, the charge density provides a more accessible and potentially experimentally verifiable parameter, especially in high-temperature regimes.

In summary, our study highlights the significant impact of breaking particle-hole symmetry on the MIT in the Hubbard model at finite temperatures. The results underscore the need to consider symmetry effects in theoretical predictions and experimental observations of phase transitions in strongly correlated electron systems.

\section{Acknowledgements}
\label{sec:acknowledgements}

The authors thank Maria Carolina Aguiar, Eduardo Miranda and Veronica Vildosola for the extremely helpful discussions and exchange of ideas.
This work was supported by the Brazilian funding agencies
CNPq (Grant Nos. 308351/2017-7, 423137/2018-2,  309789/2020-6, 313059/2020-9, and 312622/2023-6),
FAPERJ (Grant E-26/202.882/2018), FAPESP (Grant Nos. 2017/02317-2, and 2022/15453-0) and CAPES/INCT (Grant Nos. 88887.169785/2018-00, 421701/2017-0).


\vspace{0.3cm}


\end{document}